\documentclass[12pt]{iopart}

\newcommand{\be}{\begin{equation}}
\newcommand{\ee}{\end{equation}}
\newcommand{\beq}{\begin{eqnarray}}
\newcommand{\eeq}{\end{eqnarray}}

\def\q1{{q^{-1}}}
\def\dj{{{\cal D}^{(q)}_x}}

\usepackage{iopams}
\begin{document}

\title[]{Deformed quantum mechanics and $q$-Hermitian operators}

\author{A Lavagno}

\address{Dipartimento di Fisica, Politecnico di Torino, I-10129 Torino, Italy \\
Istituto Nazionale di Fisica Nucleare, Sezione di Torino, I-10126
Torino, Italy}

\ead{andrea.lavagno@polito.it}

\begin{abstract}
Starting on the basis of the non-commutative $q$-differential
calculus, we introduce a generalized $q$-deformed Schr\"odinger
equation. It can be viewed as the quantum stochastic counterpart of
a generalized classical kinetic equation, which reproduces at the
equilibrium the well-known $q$-deformed exponential stationary
distribution. In this framework, $q$-deformed adjoint of an operator
and $q$-hermitian operator properties occur in a natural way in
order to satisfy the basic quantum mechanics assumptions.
\end{abstract}



\section{Introduction}

In the recent past, there has been a great deal of interest in the
study of quantum algebra and quantum groups in connection between
several physical fields \cite{baxter}. From the seminal work of
Biedenharn \cite{bie} and Macfarlane \cite{mac}, it was clear that
the $q$-calculus, originally introduced in the study of the basic
hypergeometric series \cite{jack,gasper,Exton}, plays a central role
in the representation of the quantum groups with a deep physical
meaning and not merely a mathematical exercise. Many physical
applications have been investigated on the basis of the
$q$-deformation of the Heisenberg algebra
\cite{wess,celeghini,cerchai,meljanac,Finkelstein}. In
Ref.\cite{lava1,lava2} it was shown that a natural realization of
quantum thermostatistics of $q$-deformed bosons and fermions can be
built on the formalism of $q$-calculus. In Ref.\cite{noi1}, a
$q$-deformed Poisson bracket, invariant under the action of the
$q$-symplectic group, has been derived and a classical $q$-deformed
thermostatistics has been proposed in Ref.\cite{noi2}. Furthermore,
it is remarkable to observe that $q$-calculus is very well suited
for to describe fractal and multifractal systems. As soon as the
system exhibits a discrete-scale invariance, the natural tool is
provided by Jackson $q$-derivative and $q$-integral, which
constitute the natural generalization of the regular derivative and
integral for discretely self-similar systems \cite{erzan1}.

In the past, the study of generalized linear and non-linear
Schr\"odinger equations has attracted a lot of interest because of
many collective effects in quantum many-body models can be described
by means of effective theories with generalized one-particle
Schr\"odinger equation \cite{pita,gross,doebner,suther}. On the
other hand, it is relevant to mention that in the last years many
investigations in literature has been devoted to non-Hermitian and
pseudo-Hermitian quantum mechanics \cite{mosta1,
bender1,mosta2,bender2,jones1,bender3}.

In the framework of the $q$-Heisenberg algebra, a $q$-deformed
Schr\"odinger equations have been proposed \cite{Zhang,Micu}.
Although the proposed quantum dynamics is based on the
noncummutative differential structure on configuration space, we
believe that a fully consistent $q$-deformed formalism of the
quantum dynamics, based on the properties of the $q$-calculus, has
been still lacking.

In this paper, starting on a generalized classical kinetic equation
reproducing as stationary distribution of the well-know
$q$-exponential function, we study a generalization of the quantum
dynamics consistently with the prescriptions of the $q$-differential
calculus. At this scope, we introduce a $q$-deformed Schr\"odinger
equation with a deformed Hamiltonian which is a non-Hermitian
operator with respect to the standard (undeformed) operators
properties but its dynamics satisfies the basic assumptions of the
quantum mechanics under generalized operators properties, such as
the definition of $q$-adjoint and $q$-hermitian operator.

\section{Noncommutative differential calculus}
We shall briefly review the main features of the noncommutative
differential $q$-calculus for real numbers. It is based on the
following $q$-commutative relation among the operators $\hat x$ and
$\hat\partial_x$,
\begin{eqnarray}
\hat\partial_x\hat x=1+q\,\hat x\,\hat\partial_x \ ,\label{lei4}
\end{eqnarray}
with $q$ a real and positive parameter.\\ A realization of the above
algebra in terms of ordinary real numbers can be accomplished by the
replacement \cite{noi1,Ubriaco}
\begin{eqnarray}
&&\hat x\to x \ ,\label{rap1}\\
&&\hat\partial_x\to\dj \ ,\label{rap2}
\end{eqnarray}
where $\dj$ is the Jackson derivative \cite{jack} defined as
\begin{equation}
\dj=\frac{D^{(q)}_x-1}{(q-1)\,x} \ ,
\end{equation}
where
\begin{equation}
D^{(q)}_x=q^{x\,\partial_x} \label{dila}
\end{equation}
is the dilatation operator. Its action on an arbitrary real function
$f(x)$ is given by
\begin{equation}
\dj\,f(x)=\frac{f(q\,x)-f(x)}{(q-1)\,x} \ .
\end{equation}
The Jackson derivative satisfies some simple proprieties which will
be useful in the following. For instance, its action on a monomial
$f(x)=x^n$ is given by
\begin{equation}
\dj\,x^n=[n]_q\,x^{n-1} \ ,\label{7}
\end{equation}
and
\begin{equation}
\dj\,x^{-n}= -{[n]_q\over q^n}\,{1\over x^{n+1}} \ ,\label{7bis}
\end{equation}
where $n\geq0$ and
\begin{equation}
[n]_q=\frac{q^n-1}{q-1} \ ,\label{basic}
\end{equation}
are the so-called {\em basic}-numbers.  Moreover, we can easily
verify the following $q$-version of the Leibnitz rule
\begin{eqnarray}
\nonumber \dj\Big(f(x)\,g(x)\Big)&=&\dj\,f(x)\,g(x)+f(q\,x)\,\dj\,g(x) \ ,\\
&=&\dj\,f(x)\,g(q\,x)+f(x)\,\dj\,g(x) \ .\label{leib}
\end{eqnarray}

A relevant role in the $q$-algebra, as developed by Jackson, is
given by the {\em basic}-binomial series defined by
\begin{eqnarray}
\nonumber
(x+y)^{(n)}&=&(x+y)\,(x+q\,y)\,(x+q^2\,y)\ldots(x+q^{n-1}\,y)\\
&\equiv&\sum_{r=0}^n\Big[^{_{\displaystyle n}}_{^{\displaystyle
r}}\Big]_q\,q^{r\,(r-1)/2}\,x^{n-r}\,y^r \ ,\label{qbin}
\end{eqnarray}
where
\begin{equation}
\Big[^{_{\displaystyle n}}_{^{\displaystyle
r}}\Big]_q=\frac{[n]_q!}{[r]_q!\,[n-r]_q!} \ ,\label{qbin1}
\end{equation}
is known as the $q$-binomial coefficient which reduces to the
ordinary binomial coefficient in the $q\to1$ limit \cite{Exton}. We
should remark that Eq.(\ref{qbin1}) holds for $0\leq r\leq n$, while
it is assumed to vanish otherwise and we have defined
$[n]_q!=[n]_q\,[n-1]_q\ldots[1]_q$. Remarkably, a $q$-analogue of
the Taylor expansion has been introduced in Ref. \cite{jack} by
means of a {\em basic}-binomial (\ref{qbin}) as
\begin{eqnarray}
\hspace{-10mm}f(x)=f(a)+{(x-a)^{(1)}\over[1]!}\,\dj\,f(x)\Big|_{x=a}+
{(x-a)^{(2)}\over[2]!}\,\dj^2\,f(x)\Big|_{x=a}+\ldots
\ ,\label{Taylor}
\end{eqnarray}
where $\dj^2\equiv\dj\,\dj$ and so on.\\
Consistently with the $q$-calculus, we also introduce the {\em
basic}-integration
\begin{equation}
\int\limits_0\limits^{\lambda_0}f(x)\,d_qx=\sum_{n=0}^\infty
\Delta_q \lambda_n\,f(\lambda_n) \ ,\label{qint1}
\end{equation}
where $\Delta_q \lambda_n=\lambda_n-\lambda_{n+1}$ and
$\lambda_n=\lambda_0\,q^n$ for $0<q<1$ whilst $\Delta_q
\lambda_n=\lambda_{n-1}-\lambda_n$ and
$\lambda_n=\lambda_0\,q^{-n-1}$ for $q>1$
\cite{gasper,Exton,noi2,erzan1}. Clearly, Eq.(\ref{qint1}) is
reminiscent of the Riemann quadrature formula performed now in a
$q$-nonuniform hierarchical lattice with a variable step $\Delta_q
\lambda_n$. It is trivial to verify that
\begin{equation}
\dj\int\limits_0\limits^x f(y)\,d_qy=f(x) \ ,
\end{equation}
for any $q>0$.

Let us now introduce the following $q$-deformed exponential function
defined by the series
\begin{equation}
{\rm E}_q(x)= \sum_{k=0}^{\infty}\, \frac{x^k}{[k]_q!}=1+x+
\frac{x^2}{[2]_q!}+\frac{x^3}{[3]_q!}+\cdots \ ,\label{qexp}
\end{equation}
which will play an important role in the framework we are
introducing. The function (\ref{qexp}) defines the {\em
basic}-exponential, well known in the literature since a long time
ago, originally introduced in the study of basic hypergeometric
series \cite{gasper,Exton}. In this context, let us observe that
definition (\ref{qexp}) is fully consistent with
its Taylor expansion, as given by Eq.(\ref{Taylor}).\\
The {\em basic}-exponential is a monotonically increasing function,
$d\,{\rm E}_q(x)/dx>0$, convex, $d^2{\rm E}_q(x)/dx^2>0$, with ${\rm
E}_q(0)=1$ and reducing to the ordinary exponential in the $q\to1$
limit: ${\rm E}_1(x)\equiv \exp(x)$. An important property satisfied
by the $q$-exponential can be written formally as \cite{Exton}
\begin{equation}
{\rm E}_q(x+y)={\rm E}_q(x)\,{\rm E}_\q1(y) \ ,\label{com1}
\end{equation}
where the left hand side of Eq.(\ref{com1}) must be considered by
means of its series expansion in terms of {\em basic}-binomials:
\begin{equation}
{\rm E}_q(x+y)=\sum_{k=0}^\infty{(x+y)^{(k)}\over[k]!} \
.\label{expq1}
\end{equation}
By observing that $(x-x)^{(k)}=0$ for any $k>0$, since
$(x-x)^{(0)}=1$, from Eq.(\ref{com1}) we can see that \cite{noi2}
\begin{equation}
{\rm E}_q(x)\,{\rm E}_\q1 (-x)=1 \ .\label{inv}
\end{equation}
The above property will be crucial in the following introduction to
a consistent $q$-deformed quantum mechanics.

Among many properties, it is important to recall the following
relation \cite{Exton}
\begin{equation}
\dj {\rm E}_q(a\,x)= a\,{\rm E}_q(a\,x) \ ,\label{JDE}
\end{equation}
and its dual
\begin{equation}
\int\limits_0\limits^x{\rm E}_q(a\,y)\,d_qy={1\over a}\Big[{\rm
E}_q(a\,x)-1\Big] \ .\label{dae}
\end{equation}
Finally, it should be pointed out that Eqs.(\ref{JDE}) and
(\ref{dae}) are two important properties of the {\em
basic}-exponential which turns out to be not true if we employ the
ordinary derivative or integral.

\section{Classical $q$-deformed kinetic equation}
Starting from the realization of the $q$-algebra, defined in
Eq.s(\ref{rap1})-(\ref{rap2}), we can write for homogeneous system
the following $q$-deformed Fokker-Planck equation \cite{epjb}
\begin{equation}
\frac{\partial f_q(x,t)}{\partial t}=\dj \Big[-J_1^{(q)}(x) +
J_2^{(q)} \, \dj \Big ]\,  f_q(x,t) \; , \label{fokker}
\end{equation}
where $J_1^{(q)}(x)$ and $J_2^{(q)}$ are the drift and diffusion
coefficients, respectively.

The above equation has stationary solution $f^{(q)}_{\rm st}(x)$
that can be written as
\begin{equation}
f^{(q)}_{\rm st}(x)=N_q \, {\rm E}_q[-\Phi_q(x)]\; ,
\end{equation}
where $N_q$ is a normalization constant, $E_q[x]$ is the
$q$-deformed exponential function defined in Eq.(\ref{qexp}) and we
have defined \footnote{In the following, for simplicity, we limit
ourselves to consider the drift coefficient as a monomial function
of $x$.}
\begin{equation}
\Phi_q(x)=-\frac{1}{J_2^{(q)}}\, \int_0^x\, J_1^{(q)}(y)\, d_qy \, .
\label{phi}
\end{equation}

If we postulate a generalized Brownian motion in a $q$-deformed
classical dynamics by mean the following definition of the drift and
diffusion coefficients
\begin{equation}
J_1^{(q)}(x)= -\, \gamma \, x \; \left (q \,D^{(q)}_x+1\right ) \; ,
\ \ \ \ \ J_2^{(q)}=\gamma/\alpha \; ,
\end{equation}
where $\gamma$ is the friction constant, $\alpha$ is a constant
depending on the system and $D^{(q)}_x$ is the dilatation operator
(\ref{dila}), the stationary solution $f^{(q)}_{\rm st}(x)$ of the
above Fokker-Planck equation can be obtained as solution of the
following stationary $q$-differential equation
\begin{equation}
\dj f^{(q)}_{\rm st}(x)=-\alpha\,x \,[q\, f^{(q)}_{\rm st}(q
x)+f^{(q)}_{\rm st}(x)] \; .
\end{equation}
It easy to show that the solution of the above equation can be
written as
\begin{equation}
f^{(q)}_{\rm st}(x)=N_q \, {\rm E}_q\left[-\alpha \, x^2\right ] \;
.
\end{equation}

\section{$q$-deformed Schr\"odinger equation}

We are now able to derive a $q$-deformed Schr\"odinger equation by
means of a stochastic quantization method \cite{risken}.

Starting from the following transformation of the probability
density \be f_q(x,t)={\rm E}_q\left[-\frac{\Phi_q(x)}{2}\right ] \,
\psi_q(x,t) \, ,\ee where $\Phi_q(x)$ is the function defined in
Eq.(\ref{phi}), the $q$-deformed Fokker-Planck equation
(\ref{fokker}) can be written as \be \frac{\partial
\psi_q(x,t)}{\partial t}=J^{(q)}_2 \, \dj^2 \psi_q(x,t)-V_q(x) \,
\psi_q(x,t)\, , \ee where \be V_q(x)=\left\{\frac{1}{2}\, \dj
 J^{(q)}_1(x)+\frac{[J^{(q)}_1(x)]^2}{4\, J^{(q)}_2}\right \} \, . \ee

The above equation has the same structure of the time dependent
Schr\"odinger equation. In fact, the stochastic quantization of the
Eq.(\ref{fokker}) can be realized with the transformations
\beq &&t\rightarrow \frac{t}{-i\hbar} \, , \\
&&J^{(q)}_2\rightarrow \frac{\hbar^2}{2\,m } \, , \eeq getting the
$q$-generalized Schr\"odinger equation \be i\hbar \frac{\partial
\psi_q(x,t)}{\partial t}=H_q \, \psi_q(x,t) \, ,\label{qschro} \ee
where \be H_q=-\frac{\hbar^2}{2m}\, {{\cal D}_x^{(q)}}^2+V_q(x) \,
, \label{qhamil} \ee is the $q$-deformed Hamiltonian. Let us note
that the Hamiltonian (\ref{qhamil}) is a not-Hermitian operator
with respect to the standard definition based on the ordinary
(undeformed) scalar product of square-integrable functions
\cite{cerchai,noi1}. In the following section, we will see as this
aspect can be overridden by means the introduction of a
$q$-deformed scalar product and generalized properties of
operators inspired to $q$-calculus.

The above equation admits factorized solution
$\psi_q(x,t)=\phi(t)\,\varphi_q(x)$, where $\phi(t)$ satisfies to
the equation
\begin{equation}
i \hbar\frac{d \phi(t)}{d t}=E\phi(t) \, , \label{phit}
\end{equation}
with the standard (undeformed) solution
\begin{equation}
\phi(t)=\exp\left(-\frac{i}{\hbar}\,E\,t\right) \, ,
\end{equation}
while $\varphi_q(x)$ is the solution of time-independent
$q$-Schr\"odinger equation

\begin{equation}
H_q\, \varphi_q(x)=E\varphi_q(x) \, . \label{stationary}
\end{equation}

In one dimensional case, for a free particle ($V_q=0$) described by
the wave function $\varphi_q^f(x)$, Eq.(\ref{stationary}) becomes
\begin{equation}
\dj^2 \varphi_q^f(x)+k^2 \varphi_q^f(x)=0 \, ,
\end{equation}
where $k=\sqrt{2mE/\hbar^2}$. The solution of the previous equation
can be written as
\begin{equation}
\varphi_q^f(x)=N\, {\rm E}_q(ikx)\, . \label{plane}
\end{equation}
The above equation generalizes the plane wave function in the
framework of the $q$-calculus.

\section{$q$-deformed products and $q$-Hermitian operators}
In order to develop a consistent deformed quantum dynamics, we have
to generalize the products between functions and properties of the
operators in the framework of the $q$-calculus.

Let us start on the basis of Eq.(\ref{inv}), which implies
\begin{eqnarray}
{\rm E}_q(ix)\,\left({\rm E}_\q1(ix)\right)^\star=1
\, , \\
{\rm E}_q(ix)^\star=\left({\rm E}_\q1(ix)\right)^{-1} \, ,
\end{eqnarray}
and in terms of the $q$-plane wave (\ref{plane})
\begin{equation}
\varphi_\q1^f(x)^\star \, \varphi_q^f(x)= N^2\, .
\end{equation}
Inspired to the above equation, it appears natural to introduce the
complex $q$-conjugation of a function as
\begin{equation}
\psi_q^\dag(x)=\psi_\q1^\star(x) \, ,
\end{equation}
and, consequently, the probability density of a single particle in a
finite space as
\begin{equation}
\rho_q(x,t)=\vert \psi_q(x,t)\vert^2_q=\psi^\dag_q(x,t) \,
\psi_q(x,t)\equiv \psi^*_\q1(x,t) \, \psi_q(x,t) \, . \label{densp}
\end{equation}
Thus, the wave functions must be $q$-square-integrable functions of
con\-figuration space, that is to say the functions $\psi_q(x)$ such
that the integral \be\int \vert \psi_q(x)\vert^2_q \, d_q x\, , \ee
converges.

The function space define above it is a linear space. If $\psi_q$
and $\varphi_q$ are two $q$-square-integrable functions, any linear
combinations $\alpha\psi_q+\beta\varphi_q$, where $\alpha$ and
$\beta$ are arbitrarily chosen complex numbers, are also
$q$-square-integrable functions.

Following this line, it is possible to define a $q$-scalar product
of the function $\psi$ by the function $\varphi$ as
\begin{equation}
\langle \varphi,\psi\rangle_q=\int \varphi^\dag_q(x)\,\psi_q(x)\,
d_qx\equiv \int \varphi^\star_\q1(x)\,\psi_q(x)\, d_qx \, .
\end{equation}
This is linear with respect to $\psi$, the norm of a function
$\psi_q$ is a real, non-negative number: $\langle
\psi,\psi\rangle_q\ge 0$ and
\begin{equation}
\langle \psi,\varphi\rangle_q=\langle \varphi,\psi\rangle_q^\dag \,
.
\end{equation}
Analogously to the undeformed case, it is easy to see that from the
above properties of the $q$-scalar product follows the $q$-Schwarz
inequality \be \vert \langle \varphi,\psi\rangle_q \vert^2_q \le
\langle \varphi,\varphi\rangle_q\, \langle \psi,\psi\rangle_q \,
.\ee

Consistently with the above definitions, the $q$-adjoint of an
operator $A_q$ is defined by means of the relation
\begin{equation}
\langle \psi,A_q^\dag\varphi\rangle_q=\langle \varphi,A_q
\psi\rangle_q^\dag \, ,
\end{equation}
and, by definition, a linear operator is $q$-Hermitian if it is its
own $q$-adjoint. More explicitly, an operator $A_q$ is $q$-Hermitian
if for any two states $\varphi_q$ and $\psi_q$ we have
\begin{equation}
\langle \varphi,A_q \psi\rangle_q=\langle A_q\varphi, \psi\rangle_q
\, .\label{hermi}
\end{equation}

First of all, the above properties are crucial to have a consistent
conservation in time of the probability densities, defined in
Eq.(\ref{densp}). In fact, by taking the complex $q$-conjugation of
Eq.(\ref{qschro}), summing and integrating term by term the two
equations, we get
\begin{equation}
i\hbar \frac{\partial}{\partial t}\int \psi^\dag_q \, \psi_q \,
d_qx=\int \left[\psi^\star_\q1 (H_q\psi_q)-(H_\q1
\psi^\star_\q1)\psi_q \right ] \, d_qx =0\, ,
\end{equation}
where the last equivalence follows from the fact that the operator
Hamiltonian is $q$-Hermitian. In this context, it is relevant to
observe that it is possible to verify the above property by using
the time-spatial factorization solution
$\psi_q(x,t)=\phi(t)\,\varphi_q(x)$ of the $q$-Schr\"odinger
equation. In fact, we have
\begin{equation}
i\hbar \frac{\partial}{\partial t}\int \psi^\dag_q \, \psi_q \,
d_qx=\int \phi^\star \phi \left[\varphi^\star_\q1 (H_q
\varphi_q)-(H_\q1\varphi^\star_\q1)\varphi_q \right ] \, d_qx \, .
\label{qfatt}\end{equation} From the stationary Schr\"odinger
equation (\ref{stationary}) and its complex $q$-conjugation we have
directly
\begin{equation}
\varphi^\star_\q1 (H_q \varphi_q)=(H_\q1\varphi^\star_\q1)\varphi_q
\, ,
\end{equation}
and the terms in the square bracket of Eq.(\ref{qfatt}) goes to
zero.

\section{Observables in $q$-deformed quantum mechanics}
On the basis of the above properties, we have the recipe to
generalize the definition of observables in the framework of
$q$-deformed theory by postulating that:
\begin{itemize}
\item
with the dynamical variable $A(x,p)$ is associate the linear
operator $A_q(x,-i\hbar {\cal D}_x^{(q)})$;
\item
the mean value of this dynamical variable, when the system is in the
dynamical (normalized) state $\psi_q$, is
\begin{equation}
\langle A\rangle_q=\int \psi_q^\dag\, A_q\,\psi_q \,\,d_qx\equiv
\int \psi^\star_\q1\, A_q\,\psi_q \,\,d_qx\, . \label{meanv}
\end{equation}
\end{itemize}

Observables are real quantities, hence the expectation value
(\ref{meanv}) must be real for any state $\psi_q$:
\begin{equation}
\int \psi_q^\dag\,A_q\,\psi_q \, d_qx=\int (A_q\,\psi_q)^\dag
\,\psi_q\, d_qx \, ,
\end{equation}
therefore, on the basis of Eq.(\ref{hermi}), observables must be
represented by $q$-Hermitian operators.

If we require there is a state $\psi_q$ for which the result of
measuring the observable $A$ is unique, in other words that the
fluctuations \be (\Delta A_q)^2=\int \psi_q^\dag\, (A_q-\langle
A\rangle_q)^2 \, \psi_q \, d_qx \, ,\ee must vanish, we obtain the
following $q$-eigenvalue equation of a $q$-Hermitian operator $A_q$
with eigenvalue $a$
\begin{equation}
A_q \,\varphi_q= a \,\varphi_q  \, .
\end{equation}
As a consequence, the eigenvalues of a $q$-Hermitian operator are
real because $\langle A\rangle_q$ is real for any state; in
particular for an eigenstate with the eigenvalue $a$ for which
$\langle A\rangle_q=a$.

Furthermore, as in the undeformed case, two eigenfunctions
$\psi_{q,1}$ and $\psi_{q,2}$ of the $q$-Hermitian operator $A_q$,
corresponding to different eigenvalues $a_1$ and $a_2$, are
orthogonal. We can always normalize the eigenfunction, therefore we
can chose all the eigenvalues of a $q$-Hermitian operator
orthonormal,  i.e. \be \int \psi_{q,n}^\dag\,\psi_{q,m} \,
d_qx=\delta_{n,m} \, . \ee Consequently, two eigenfunctions
$\psi_{q,1}$ and $\psi_{q,2}$ belonging to different eigenvalues are
linearly independent.

It easy to see that, adapting step by step the undeformed case to
the introduced $q$-deformed framework, the totality of the linearly
independent eigenfunctions $\{\psi_{q,n}\}$ of $q$-Hermitian
operator $A_q$ form a complete (orthonormal) set in the space of the
previously introduced $q$-square-integrable functions. In other
words, if $\psi_q$ is any state of a system, then it can be expanded
in terms of the eigenfunctions (with a discrete spectrum) of the
corresponding $q$-Hermitian operator $A_q$ associate to the
observable: \be \psi_q=\sum_n c_{q,n}\, \psi_{q,n} \, , \ee where
\be c_{q,n}=\int \psi_{q,n}^\dag\,\psi_q\, d_qx \, . \ee

The above expansion allows us, as usual, to write the expectation
value of $A_q$ in the normed state $\psi_q$ as \be \langle
A\rangle_q=\int \psi_q^\dag\, A_q\,\psi_q \,\,d_qx= \sum_n \vert
c_{q,n}\vert^2_q \, a_n \, , \ee where $\{a_n\}$ are the set of
eigenvalues (assumed, for simplicity, discrete and non-degenerate)
and the normalization condition of the wave function can be written
in the form \be \sum_n \vert c_{q,n}\vert^2_q=1 \, .\ee

\section{Conclusions}
On the basis of the stochastic quantization procedure and on the
$q$-differential calculus, we have obtained a generalized linear
Schr\"odinger equation which involves a $q$-deformed Hamiltonian
that is non-Hermitian with respect to the standard (undeformed)
definition. However, under an appropriate generalization of the
operators properties and the introduction of a $q$-deformed scalar
product in the space of $q$-square-integral wave functions, such
equation of motion satisfies the basic quantum mechanics
assumptions.

Although a complete physical and mathematical description of the
introduced quantum dynamical equations lies out the scope of this
paper, we think that the results derived here appear to provide a
deeper insight into a full consistent $q$-deformed quantum mechanics
in the framework of the $q$-calculus and may be a relevant starting
point for future investigations.

\vspace{0.6cm}
\noindent {\bf Acknowledgment}\\
\\
It is a pleasure to thank A M Scarfone for useful discussions.

\end{document}